\RequirePackage{fix-cm}
\documentclass[smallextended]{svjour3}       
\smartqed  
\usepackage{graphicx}
\newcommand {\bm}[1]{\mathbf#1}
\newcommand {\lesssim}{\ \raise.3ex\hbox{$<$}\kern-0.8em\lower.7ex\hbox{$\sim$}\ }
\newcommand {\gesim}{\ \raise.3ex\hbox{$>$}\kern-0.8em\lower.7ex\hbox{$\sim$}\ }

\newcommand \beq{\begin{eqnarray}}
\newcommand \eeq{\end{eqnarray}}

\begin{document}

\title{Multi-body correlations in SU(3) Fermi gases
}


\author{H. Tajima         \and
        P. Naidon 
}


\institute{H. Tajima \at
              Quantum Hadron Physics Laboratory, RIKEN Nishina Center, Hirosawa 2-1, Wako, Japan \\
              Tel.: +81 048-467-9343\\
              Fax: +81 048-462-4698\\
              \email{hiroyuki.ajima@riken.jp}           
           \and
            P. Naidon \at
              Quantum Hadron Physics Laboratory, RIKEN Nishina Center, Hirosawa 2-1, Wako, Japan
}

\date{Received: date / Accepted: date}

\maketitle

\begin{abstract}
We investigate strong-coupling effects in a three-component atomic Fermi gas. 
It is a promising candidate for simulating quantum chromodynamics (QCD), and  
furthermore, the emergence of various phenomena such as color superfluidity and Efimov effect are anticipated in this system.
In this paper, we study the effects of two-body and three-body correlations by means of the many-body $T$-matrix approximation (TMA)
as well as the  Skorniakov-Ter-Martirosian (STM) equation with medium corrections.
We investigate the effects of finite temperature and chemical potential on  the trimer binding energy at the superfluid critical point of the unitarity limit.

\end{abstract}

\section{Introduction}
\label{intro}
Ultracold atomic gases give us ideal testing grounds for the study of various strongly correlated quantum systems \cite{Bloch,Giorgini}.
The controllability of physical parameters such as interatomic interactions enables us to use these atomic systems as quantum simulators for other systems, ranging from high-$T_{\rm c}$ superconductors \cite{Hofstetter,Chin,Gross} to neutron star matter \cite{Gezerlis,Navon,Horikoshi,Tajima,vanWyk}. 
In particular, a three-component Fermi gas is expected to be analogous to quantum chromodynamics (QCD) \cite{Fukushima} where quarks with three colors strongly interact with each other. 
The crossover from a trimer phase to a color superfluid phase \cite{He,Paananen,Ozawa} has been theoretically proposed in this system \cite{Floerchinger,Nishida,Kirk}
in analogy with the hadron phase and color superconducting phase of QCD.
Conventional superfluids have already been realized in two-component Fermi gases of $^{40}$K \cite{Regal} and $^6$Li \cite{Zwierlein} atoms and have been extensively discussed \cite{Bloch,Giorgini}.
In the case of three components, Fermi degeneracy has been achieved experimentally \cite{Ottenstein,Ohara} and
the existence of three-body bound states called the Efimov trimers \cite{Efimov,Naidon,Naidon2,Naidon3} has been experimentally confirmed \cite{Williams,Wenz,Nakajima}.
However, neither the color superfluidity nor Efimov trimer phase have been realized in current experiments yet.
The study of these many-body states constitutes a great challenge for understanding strong-coupling effects in both cold atom systems and dense QCD matter.
\par
In this paper, we theoretically investigate two-body and three-body correlations in a symmetric three-component Fermi gas.
By using the many-body $T$-matrix approximation (TMA) \cite{Perali,Tsuchiya,Tajima2}, which successfully describes the crossover from weak-coupling Bardeen-Cooper-Schrieffer (BCS) Fermi superfluidity to the Bose-Einstein condensation (BEC) of molecules in two-component Fermi gases, we first incorporate effects of superfluid fluctuations associated with two-body correlations.
Specifically, we consider a two-channel model that physically describes a narrow resonance with finite negative effective range \cite{Nishida}.
We calculate the superfluid phase transition temperature $T_{\rm c}$ and critical chemical potential $\mu_{\rm c}$ as a function of the effective range $r_{\rm e}$ where the scattering length $a$ diverges.
We then investigate effects of the medium on the trimer binding energy $E_3$ by means of the Skorniakov-Ter-Martirosian (STM) equation \cite{STM} with medium corrections, where the STM equation is known as an exact equation to depict Efimov physics in the three-body problem \cite{Naidon3}. 
In the following, we use $\hbar=k_{\rm B}=1$ and the system volume is taken to be unity, for simplicity.

\section{Formulation}
\label{formulation}
We start from the two-channel Hamiltonian for three-component symmetric fermions given by
\beq
\label{eq1}
H&=&\sum_{i=1,2,3}\sum_{\bm{p}}\xi_{\bm{p}}c_{\bm{p},i}^{\dag}c_{\bm{p},i}+\sum_{i<j}\sum_{\bm{q}}\left(\varepsilon_{\bm{q}}^{\rm B}+\nu-2\mu\right)b_{\bm{q},ij}^{\dag}b_{\bm{q},ij} \cr
&&+g\sum_{i<j}\sum_{\bm{p},\bm{q}}\left(b_{\bm{q},ij}^{\dag}c_{-\bm{p},i}c_{\bm{p}+\bm{q},j}+ {\rm H. c.}\right),
\eeq
where $\xi_{\bm{p}}=\bm{p}^2/2m-\mu$ and $\varepsilon_{\bm{q}}^{\rm B}=\bm{q}^2/4m$ are the kinetic energies of a Fermi atom with mass $m$ measured from the chemical potential $\mu$ and a diatomic molecules, respectively ($\bm{p}$ and $\bm{q}$ are the momenta).
$c_{\bm{p},i}$ $(i=1,2,3)$ and $b_{\bm{q},ij}$ $(i<j)$ are the annihilation operators of a Fermi atom with the hyperfine state $i$ and a diatomic molecule of $i$-$j$ pair, respectively.  
In our model, the threshold energy of a diatomic molecule $\nu$ and the Feshbach coupling $g$ can be written in terms of the scattering length $a$ and effective range $r_{\rm e}$ as follows,
\beq
\label{eq2}
\frac{1}{a}=\frac{1}{2}r_{\rm e}\nu+\frac{2}{\pi}\Lambda, \quad r_{\rm e}=-\frac{8\pi}{m^2g^2},
\eeq 
where $\Lambda$ is the ultraviolet momentum cutoff. In this paper, we focus on $1/a=0$.
\par
We calculate the superfluid phase transition temperature $T_{\rm c}$ and chemical potential $\mu_{\rm c}$ within the framework of the many-body $T$-matrix approximation (TMA).
The atomic thermal Green's function $G_i(\bm{p},i\omega_n)$ with the fermionic Matsubara frequency $\omega_n=(2n+1)\pi T$ is given by
\beq
\label{eq3}
G_{i}(\bm{p},i\omega_n)=\frac{1}{i\omega_n-\xi_{\bm{p}}-\Sigma_{i}(\bm{p},i\omega_n)},
\eeq
\begin{figure}
\begin{center}
  \includegraphics[width=0.6\textwidth]{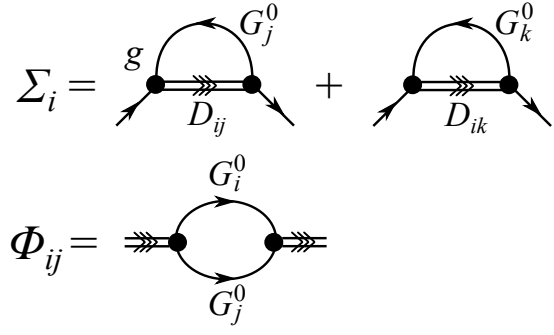}
\end{center}
\caption{Feynman diagrams describing the atomic and diatomic self-energy $\Sigma_i$ and $\Phi_{ij}$.
The solid and double-solid lines represent the bare atomic Green's function $G_{i}^{0}$ and the dressed diatomic Green's function $D_{ij}$, respectively.The black dot shows the Feshbach coupling $g$.}
\label{fig1}       
\end{figure}
The atomic self-energy $\Sigma_{i}(\bm{p},i\omega_n)$ diagrammatically shown in Fig.~\ref{fig1} is given by
\beq
\label{eq4}
\Sigma_{i}(\bm{p},i\omega_n)&=&g^2T\sum_{\bm{q},i\nu_{n'}}\left[D_{ij}(\bm{q},i\nu_{n'})G_{j}^0(\bm{q}-\bm{p},i\omega_n-i\nu_{n'})\right.\cr
&&+\left.D_{ik}(\bm{q},i\nu_{n'})G_{k}^0(\bm{q}-\bm{p},i\omega_n-i\nu_{n'})\right] \quad (i\neq j \neq k),
\eeq
where 
\beq
\label{eq5}
D_{ij}(\bm{q},i\nu_{n'})=\frac{1}{i\nu_{n'}-\varepsilon_{\bm{q}}^{\rm B}-\nu+2\mu-\Phi_{ij}(\bm{q},i\nu_{n'})}
\eeq
is the thermal Green's function of a $i$-$j$ diatomic pair ($\nu_{n'}=2n'\pi T$ is the bosonic Matsubara frequency),
which involves the self-energy $\Phi_{ij}(\bm{q},i\nu_{n'})$ (see Fig.~\ref{fig1}) given by 
\beq
\label{eq6}
\Phi_{ij}(\bm{q},i\nu_{n'})=-g^2T\sum_{\bm{p},i\omega_n}G_{i}^0(\bm{p}+\bm{q},i\omega_n+i\nu_{n'})G_{j}^0(-\bm{p},-i\omega_n).
\eeq
We note that $G_{i}^0(\bm{p},i\omega_n)=1/(i\omega_n-\xi_{\bm{p}})$ in Eqs. (\ref{eq4}) and (\ref{eq6}) is the bare atomic Green's function. $T_{\rm c}$ and $\mu_{\rm c}$ are determined by solving the particle number equation
\beq
\label{eq7}
N=T\sum_{i=1,2,3}\sum_{\bm{p},i\omega_n}G_{i}(\bm{p},i\omega_n)-2T\sum_{i<j}\sum_{\bm{q},i\nu_{n'}}D_{ij}(\bm{q},i\nu_{n'}),
\eeq
where $N$ is the total atomic number and the Thouless criterion \cite{Tajima2},
\beq
\label{eq8}
\left[D_{ij}(\bm{q}=0,i\nu_{n'}=0)\right]^{-1}=0.
\eeq
\par
After obtaining $T_{\rm c}$ and $\mu_{\rm c}$, we determine the trimer binding energy $E_3$ by solving the Skorniakov-Ter-Martirosian (STM) equation \cite{STM} in the presence of medium corrections.
In our model, the STM equation is given by
\beq
\label{eq9}
&&\left[\frac{r_{\rm e}\kappa(\bm{q})^2}{2}+4\pi\sum_{\bm{p}}\left\{\frac{F(\bm{p},\bm{q})}{\bm{p}^2+\kappa(\bm{q})^2}-\frac{1}{\bm{p}^2}\right\}\right]
\chi(\bm{q})\cr
&&=-8\pi\sum_{\bm{p}'}\frac{F(\bm{p}',\bm{q})\chi(\bm{p}'+\bm{q/2})}{\bm{p}'^2+\kappa(\bm{q})^2},
\eeq
where $\kappa(\bm{q})^2=\frac{3}{4}\bm{q}^2-E_3$.
The medium corrections are included in the statistical factor $F(\bm{p},\bm{q})$.
Considering the Pauli-blocking effect on Fermi atoms, we introduce 
\beq
\label{eq10}
F(\bm{p},\bm{q})=[1-f(\xi_{\bm{p}+{\bm{q}/2}})][1-f(\xi_{\bm{p}-{\bm{q}/2}})],
\eeq
where $f(\xi_{\bm{p}})=1/\left[e^{(\bm{p}^2/2m-\mu_{\rm c})/T_{\rm c}}+1\right]$ is the Fermi-Dirac distribution function.
Eq. (\ref{eq10}) is a generalization of Ref.~\cite{Niemann} to the finite temperature where the step functions are replaced by $f(\xi_{\bm{p}})$.
We note that it can be regarded as a particle-particle (pp) pair contribution above the Fermi sea.
In addition, we also calculate $E_{3}$ by using $F(\bm{p},\bm{q})$ including the hole-hole (hh) pair contribution below the Fermi sea, given by
\beq
\label{eq11}
F(\bm{p},\bm{q})=[1-f(\xi_{\bm{p}+{\bm{q}/2}})][1-f(\xi_{\bm{p}-{\bm{q}/2}})]-f(\xi_{\bm{p}+{\bm{q}/2}})f(\xi_{\bm{p}-{\bm{q}/2}}).
\eeq
We note that both factors go to unity in the vacuum limit $\mu \rightarrow -\infty$ and Eq. (\ref{eq9}) reduces to the ordinary STM equation for a three-body system in this limit.

\section{Results}
\label{results}
\begin{figure}
\begin{center}
  \includegraphics[width=0.65\textwidth]{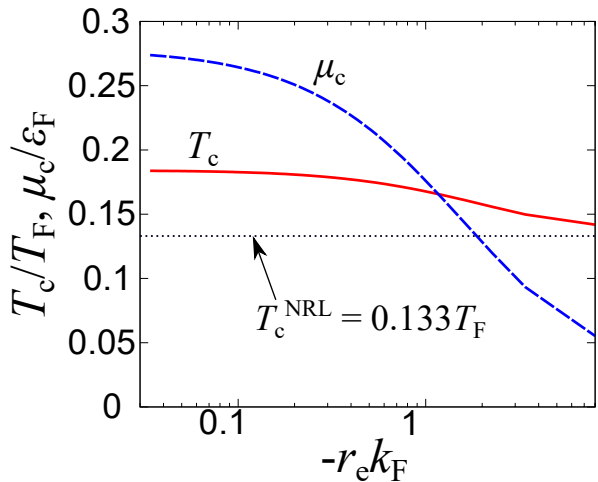}
\end{center}
\caption{Calculated superfluid phase transition temperature $T_{\rm c}$ (solid curve) and critical chemical potential $\mu_{\rm c}$ (dashed curve) as functions of the negative effective range $-r_{\rm e}k_{\rm F}$ at $1/a=0$.
The dotted line represents $T_{\rm c}^{\rm NRL}=0.133T_{\rm F}$ obtained from Eq. (\ref{eq12}).}
\label{fig2}       
\end{figure}
Figure~\ref{fig2} shows the effective-range dependence of the superfluid phase transition temperature $T_{\rm c}/T_{\rm F}$ and the critical chemical potential $\mu_{\rm c}/\varepsilon_{\rm F}$ at $1/a=0$, where $T_{\rm F}$ and $\varepsilon_{\rm F}$ are the Fermi temperature and Fermi energy, respectively.
Both quantities gradually decrease with increasing the absolute value of the effective range.
A similar behavior can be seen in a strongly interacting two-component Fermi gas with finite negative effective range \cite{Tajima2}.
In the narrow resonance limit ($g\rightarrow 0$, $r_{\rm e}\rightarrow -\infty$), since the self-energy corrections disappears in Eqs. (\ref{eq5}) and (\ref{eq8}), $\mu_{\rm c}$ goes to $\nu/2$ $(=0)$.  
Therefore, in the large-negative-effective-range region, $T_{\rm c}$ approaches $T_{\rm c}^{\rm NRL}=0.133T_{\rm F}$, which is obtained by solving 
\beq
\label{eq12}
N=3\sum_{\bm{p}}f\left(\varepsilon_{\bm{p}}\right)+6\sum_{\bm{q}}b\left(\varepsilon_{\bm{q}}^{\rm B}\right),
\eeq
where $\varepsilon_{\bm{p}}=\bm{p}^2/2m$ and 
$b(\varepsilon_{\bm{q}}^{\rm B})=1/(e^{\varepsilon_{\bm{q}}^{\rm B}/T}-1)$ is the Bose-Einstein distribution function.
Eq. (\ref{eq12}) is obtained from Eq. (\ref{eq7}) by taking limits of $\mu\rightarrow 0$ and $g\rightarrow 0$.
$T_{\rm c}^{\rm NRL}$ is close to the BEC temperature of molecules in the strong-coupling limit at zero effective range given by $T_{\rm BEC}\simeq 0.137T_{\rm F}$ \cite{Ozawa} since the system is dominated by diatomic molecules. 
We note that the small difference between $T_{\rm c}^{\rm NRL}$ and $T_{\rm BEC}$ originates from the first term of Eq. (\ref{eq12}) corresponding to the contribution of thermal-excited atoms.
\par
\begin{figure}
\begin{center}
  \includegraphics[width=0.65\textwidth]{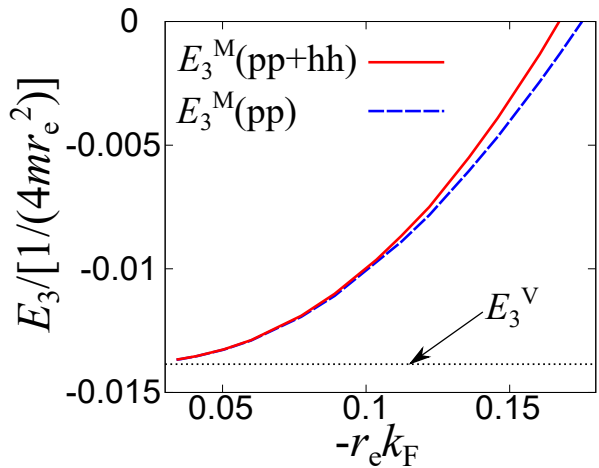}
\end{center}
\caption{Calculated Efimov trimer binding energy $E_3$ as a function of the negative effective range $-r_{\rm e}k_{\rm F}$ at $1/a=0$.
The dashed and solid curves show $E_{\rm 3}^{\rm M}$ obtained by using Eqs. (\ref{eq10}) and (\ref{eq11}), respectively.
The dotted line is the Efimov trimer binding energy in vacuum $E_{\rm 3}^{\rm V}=-0.0138542/(4mr_{\rm e}^2)$ obtained from the ordinary STM equation [Eq. (\ref{eq9}) with $F(\bm{p},\bm{q})=1$]. }
\label{fig3}       
\end{figure}

Figure \ref{fig3} shows the effective-range dependence of the binding energy of an Efimov trimer in medium $E_3^{\rm M}$ calculated by solving Eq. (\ref{eq9}) with $T_{\rm c}$ and $\mu_{\rm c}$ shown in Fig. \ref{fig2}.
The dashed and solid curves are obtained by using Eqs. (\ref{eq10}) and (\ref{eq11}) for $F(\bm{p},\bm{q})$, respectively.
In the zero effective-range limit, both curves coincide with the Efimov trimer binding energy in vacuum given by $E_{\rm 3}^{\rm V}=-0.0138542/(4mr_{\rm e}^2)$ \cite{Nishida,Naidon3} since the contribution of the high-energy region in the integral of Eq. (\ref{eq9}) is rather important there.
If one regards the horizontal axis $-r_{\rm e}k_{\rm F}$ as a measure of the particle density $N=k_{\rm F}^2/(2\pi^2)$ with fixed $r_{\rm e}$,
the limit ($r_{\rm e}k_{\rm F}\rightarrow 0_{-}$) corresponds to the low-density limit.
In this sense, this cold atomic system has a phase structure resembling dense QCD matter where all quarks are confined in hadrons in the low-density regime.
With an increasing negative effective range, medium effects suppress the binding of Efimov trimers and finally $E_{\rm 3}^{\rm M}$ disappears around $r_{\rm e}k_{\rm F}\simeq -0.17$.
However, this behavior does not necessarily mean the disappearance of the trimer states at this point.
There may still be trimer state solutions of the STM equation at a positive energy ($E_3>0$), called Cooper triple states \cite{Niemann}.
These states can be understood as a generalization of the Cooper problem, where two electrons can form a so-called Cooper pair in the presence of a Fermi surface and an infinitesimally attractive interaction \cite{BCS}.
In the case of Cooper triples, one also has to consider the Pauli-blocking effect on fermionic trimers, in contrast to Cooper pairs which are bosonic. 
To understand such a many-body state, a self-consistent treatment of two-body and three-body correlations is necessary, which is left as an interesting future work.
\par
The difference between two curves of $E_{3}^{\rm M}$ in Fig. \ref{fig3} comes from the hole-hole (hh) pair contribution described by the second term of Eq. (\ref{eq11}).
The appearance of hole-hole pair excitations at the same time as the particle-particle excitations would be natural in the presence of Fermi seas.
As shown in Fig. \ref{fig3} this effect becomes slightly larger with increasing the negative effective range.
One can find that the qualitative behavior of negative $E_3^{\rm M}$ can be captured by considering only the particle-particle (pp) pair contribution.
\par
\section{Summary}
\label{summary}
To summarize, we have theoretically investigated the effects of two-body and three-body correlations in a three-component Fermi gas.
By using the many-body $T$-matrix approximation to incorporate effects of two-body pairing fluctuations, we have numerically calculated the superfluid phase transition temperature and critical chemical potential as functions of the negative effective range.
Furthermore, we have solved the Skorniakov-Ter-Martirosian equation in the medium background.
The Efimov trimer binding is suppressed with increasing the density or negative effective range by medium corrections associated with the Pauli-blocking effects on Fermi atoms in the intermediate state.
This behavior is quite similar to the quark deconfinement, where the finite density breaks up a hadron into three quarks.
We hope that our study contributes to the understanding of this phenomenon
in both condensed matter and high-energy physics.

\begin{acknowledgements}
We thank Y. Nishida, T. Hatsuda, and G. Baym for useful discussions. 
H. T. was supported by a Grant-in-Aid for JSPS fellows (No. 17J03975).
P. N. was supported by RIKEN Incentive Research Project.
This work was partially supported by iTHEMS Program.
\end{acknowledgements}



\end{document}